# Observation of the Topologically Originated Edge States in large-gap Quasi-One-Dimensional α-Bi$_4$Br$_4$

Pengcheng Mao[1,2,†], Maoyuan Wang[1,†], Dashuai Ma[1,†], Dongyun Chen[1,2,†], Cheng-cheng Liu[1], Xiang Li[1,2], Jingchuan Zheng[1], Yongkai Li[1], Huixia Yang[1,2], Qinsheng Wang[1,2], Junxi Duan[1,2], Jie Ma[1], Yuanchang Li[1], Hailong Chen[3,4], Zhi Xu[3,4,5], Shuang Jia[6], Junfeng Han[1,2]\*, Wende Xiao[1,2]\* and Yugui Yao[1,2]\*

[1]Key laboratory of advanced optoelectronic quantum architecture and measurement, ministry of education, School of Physics, Beijing Institute of Technology, Beijing, 100081, China

[2]Micronano Centre, Beijing Key Lab of Nanophotonics & Ultrafine Optoelectronic Systems, Beijing Institute of Technology, Beijing, 100081, China

[3]Beijing National Laboratory for Condensed Matter Physics and Institute of Physics, Chinese Academy of Sciences, Beijing, 100190, China

[4]Songshan Lake Materials Laboratory, Dongguan, Guangdong, 523808, China

[5]School of Physical Sciences, University of Chinese Academy of Sciences, Beijing, 100190, China

[6]ICQE, School of Physics, Peking University, Beijing, 100871, China

**Two-dimensional topological insulator features time-reversal-invariant spin-**

___________________________________________________________________

[†]These authors contributed equally: Pengcheng Mao, Maoyuan Wang, Dashuai Ma, Dongyun Chen.

\*email: pkuhjf@bit.edu.cn, wdxiao@bit.edu.cn, ygyao@bit.edu.cn.

**momentum-locked one-dimensional (1D) edge states with a linear energy dispersion. However, experimental access to 1D edge states is still of great challenge and only limited to few techniques to date. Here, by using infrared absorption spectroscopy, we observed robust topologically originated edge states in α-Bi$_4$Br$_4$ belts with definitive signature of strong infrared absorption at belt sides and distinct anisotropy with respect to light polarizations, which is further supported by first-principles calculations. Our work demonstrates for the first time that the infrared spectroscopy can offer a power-efficient approach in experimentally probing 1D edge states of topological materials.**

Topological insulator features an insulating bulk band gap and time-reversal-invariant topological surface/edge states, offering new opportunities for unprecedented advances in technologies such as spintronics, microelectronics and infrared detectors [1-9]. The family of bismuth halogenides, including α/β-Bi$_4$Br$_4$ and α/β-Bi$_4$I$_4$, has recently been attracting great interest, and various novel properties have been reported, e.g. strong topological insulator [10], weak topological insulator [11], high-order topological insulator [12], superconductivity [13,14] and quantum spin Hall effect for single layer [15]. Among them, bulk α-Bi$_4$Br$_4$ crystallizes in a monoclinic space group *C2/m* (No. 12), and can be viewed as a parallel arrangement of one-dimensional (1D) infinite molecular chains [16]. Within each single molecular chain, one Bi atomic chain is sandwiched by two Bi/Br atomic chains from both sides, as shown in Fig. 1a. The 1D molecular chains run along the *b*-direction and are closely packed along the *a*- direction via van der Waals (vdW) interaction, forming single layers of $Bi_4Br_4$ along the *ab* plane. The mirror-reflected single layers (with the *b*-direction as a normal) are vdW stacked alternatively along the *c*-direction to form the bulk α-$Bi_4Br_4$ [17]. Figure 1b depicts the Brillouin zone of α-$Bi_4Br_4$ with all high-symmetry points labelled. The calculated band structure of the bulk α-$Bi_4Br_4$ exhibits a direct band gap of 0.145 eV near the *M* point (Fig. 1c). As the band inversion occurs twice at both *L* and *M* points, α-$Bi_4Br_4$ was classified as a $Z_2$ trivial insulator [18] according to the Fu-Kane criterion [19].

In contrast to bulk α-$Bi_4Br_4$, our previous calculations revealed that a free-standing single-layer $Bi_4Br_4$ is a quantum spin Hall insulator that hosts gapless helical edge states [15]. Such topologically protected edge states are still preserved at the single-layer step edges of the (001)-surface of α-$Bi_4Br_4$ due to the weak interlayer coupling (Fig. 1d). Moreover, for the bilayer step edges, each single-layer step edge contributes a branch of topological edge states and the coupling of the two branches of edge states leads to the opening of a tiny gap of ~ 20 meV (Fig. 1e) [20]. Despite of the single-layer and the bilayer step edges possessing nontrivial and trivial $Z_2$ topological invariant, respectively, they share a similar topological origination. In addition, the multilayer step edge that can also be regarded as a (100)-oriented facet of α-$Bi_4Br_4$ possesses surface states with a narrow band gap near the $\bar{G}$ point (Fig. 1f). Therefore, all these kinds of step edges with various step heights accommodate topologically originated edge states (TOES) in the band gap of bulk α-$Bi_4Br_4$. However, such spatially localized 1D edge states can be hardly probed in experiments except scanning tunneling microscopy (STM) in combination with angle-resolved photoemission spectroscopy (ARPES) to date [21-28]. In this work, we demonstrate that the TOES of α-Bi$_4$Br$_4$ can be spatially resolved by using infrared spectroscopy and furthermore we investigate its anisotropic properties with polarized incident light in combined with first-principles calculations.

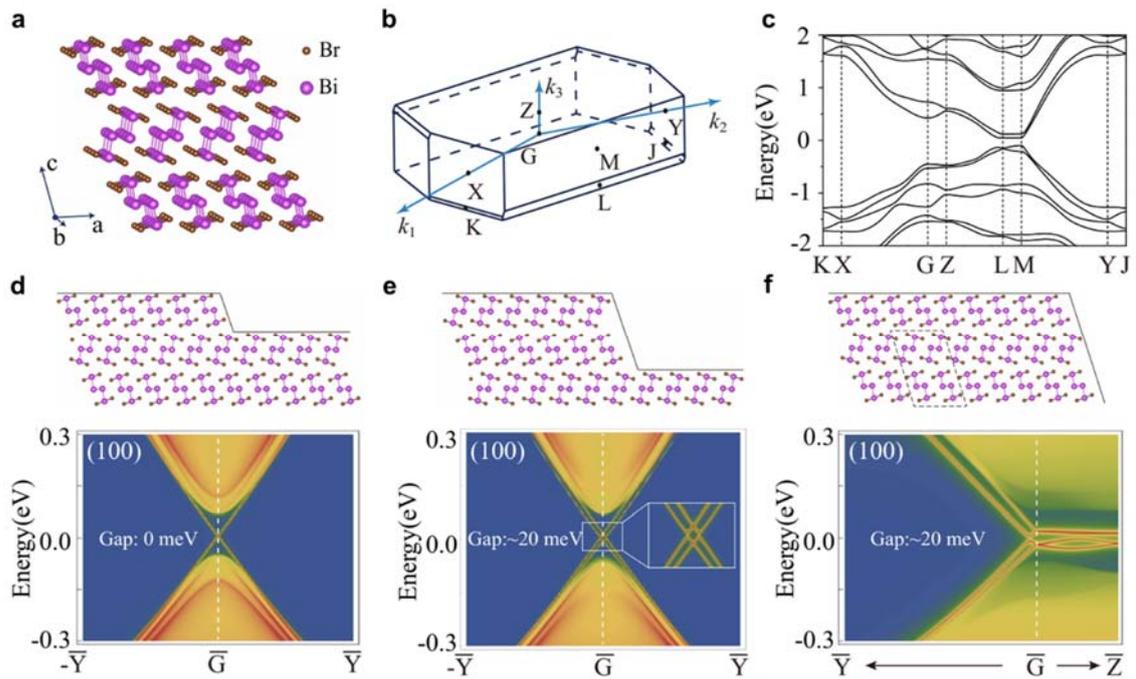

**Fig. 1 Electronic structures of α-Bi$_4$Br$_4$.** (**a**, **b**) Crystal structure and Brillouin zone of α-Bi$_4$Br$_4$. The chain runs along the *b* axis in the real space while it parallels with the **k$_1$-k$_2$** direction in the reciprocal space. (**c**) Band structure of α-Bi$_4$Br$_4$. (**d, e, f**) Upper panels, side views of the steps running along the *b* axis with single-layer, bilayer and multilayer heights on the (001) surface of α-Bi$_4$Br$_4$, respectively. Note that the multilayer step edge can also be regarded as a (100)-oriented facet along the *bc* plane of α-Bi$_4$Br$_4$. The unit cell of α-Bi$_4$Br$_4$ is highlighted in Fig. 1**f** by dashed black lines. Lower panels, band structures of the single-layer, bilayer and multilayer steps,

respectively. The gap of edge states is ~ 20 meV for bilayer and multilayer steps, while it is zero for single-layer step. The Fermi level is set to the centre of the bulk gap. $\bar{G}$, $\bar{Y}$ and $\bar{Z}$ are the projected points of $G$, $Y$ and $Z$ (Fig. 1**b**) in the surface Brillouin zone of the (100) surface.

First, we have grown high-quality α-Bi$_4$Br$_4$ single crystals by an optimized Bi self-flux method [29]. The (00$l$) reflections of X-ray diffraction (XRD) pattern (Fig. 2a), particularly the (007) and (00$\underline{11}$) odd peaks, indicate that the as-grown crystals are α-Bi$_4$Br$_4$ with (001)-oriented tops. The high-resoltuion transmision electron microscopy (TEM) images and the electron diffraction patterns reveal the high quanlity of the crystal (Fig. 2b). A Bi:Br atomic ratio of 1:1 determined by energy dispersion X-ray spectra and X-ray photoelectron spectra is consistent with the stoichiometry of α-Bi$_4$Br$_4$. Due to the weak vdW-type interlayer and interchain coupling in α-Bi$_4$Br$_4$ [15], it is easy to obtain α-Bi$_4$Br$_4$ belts with (001)-oriented flat tops along the *ab* plane and straight edges along the 1D chain axis via mechanical exfoliation, as shown in the images of optical microscopy (OM) (Fig. 2c). Figure 2d depicts a large-scale three-dimensional (3D) atomic force microscopy (AFM) image of the as-prepared α-Bi$_4$Br$_4$ belt, where zoom-in images display its flat (001)-oriented top surface with few steps (Fig. 2e) and a bunch of straight steps on the belt side (Fig. 2f). According to the theoretical analysis above, these long and straight step edges with various thickness would host TOES with zero or tiny gaps that can be well detected by infrared spectroscopy. We therefore systematically performed infrared absorption measurements on the α-Bi$_4$Br$_4$ belt, expecting a much stronger infrared absorption from the belt side than that from the top due to the former high step density.

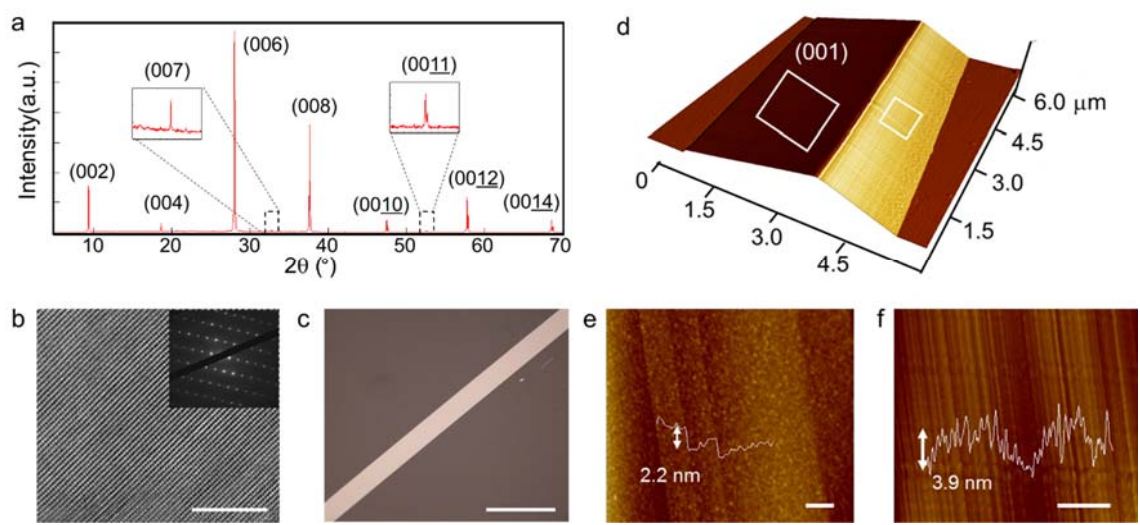

**Fig. 2 Characterization of single-crystalline α-Bi₄Br₄ belts.** (a) XRD pattern of an α-Bi₄Br₄ single crystal with (001)-oriented top. The insets show the characteristic (007) and (00$\underline{11}$) odd reflections for α-Bi₄Br₄. (b) High-resolution TEM image showing the high quality of the α-Bi₄Br₄ single crystal. Scale bar, 10 nm. The inset shows an electron diffraction pattern of the α-Bi₄Br₄ single crystal. (c) OM image of a single-crystalline α-Bi₄Br₄ belt, showing a flat top and straight edges. Scale bar, 100 μm. (d) 3D AFM image of the α-Bi₄Br₄ belt. (e, f) Zoom-in AFM images acquired at the top and side of the belt, respectively. Scale bar, 100 nm.

Infrared absorption spectra are collected along a line perpendicularly across an exfoliated α-Bi₄Br₄ belt with a width of ~20 μm and a thickness of ~805 nm on a CaF₂ substrate (Fig. 3a) [29]. Figure 3b shows the background-subtracted spectra acquired from three typical regions (indicated in Fig. 3a), namely, the centre (blue curve) and side (red curve) of the α-Bi$_4$Br$_4$ belt and the CaF$_2$ substrate (black curve): (1) the spectrum taken from the centre (blue curve) exhibits an obvious absorption slope in the region of 1,750 ~ 1,850 cm$^{-1}$. A band gap of 0.22 eV (ca. 1770 cm$^{-1}$) for bulk α-Bi$_4$Br$_4$ is estimated [29], slightly wider than the calculated result [18]. A weak infrared absorption has also been observed in the small wavenumber region (< 1,700 cm$^{-1}$), which is probably due to the absorption of the impurity/defect states or the intra-band absorption in the conduction band. (2) The infrared absorbance (OD), which is defined as OD = $log(I_0/I)$, where $I_0$ and $I$ are the intensities of the incident and transmission light, respectively, measured at the belt side (red curve) dramatically reduced with respect to that of the centre in the large wavenumber region (> 1,900 cm$^{-1}$) arising from the leakage of the incident light. However, the infrared absorption in the small wavenumber region (< 1,700 cm$^{-1}$) is surprisingly stronger than that from the centre.

In order to further shed light on the strong infrared absorption from the belt side, we image the two-dimensional (2D) infrared absorption map. In the small wavenumber region (1,000 ~ 1,700 cm$^{-1}$), the absorbance from the belt side in Fig. 3c is obviously much stronger than that from the centre. As a comparison, the absorption image taken at 2,500 ~ 3,200 cm$^{-1}$ (Fig. 3d) corresponding to the inter-band absorption of the bulk shows a stronger absorbance from the belt centre than from the side. More importantly, all sides regardless of straight (left and right sides) or irregular shapes (lower side) exhibit almost identical absorbance in the small wavenumber region, indicating that this kind of infrared absorption involved electronic states are very robust against the variation of local geometry and/or chemical bonds. Such robustness suggests that the strong infrared absorption from the sides in the small wavenumber region can be accounted for by the aforementioned TOES.

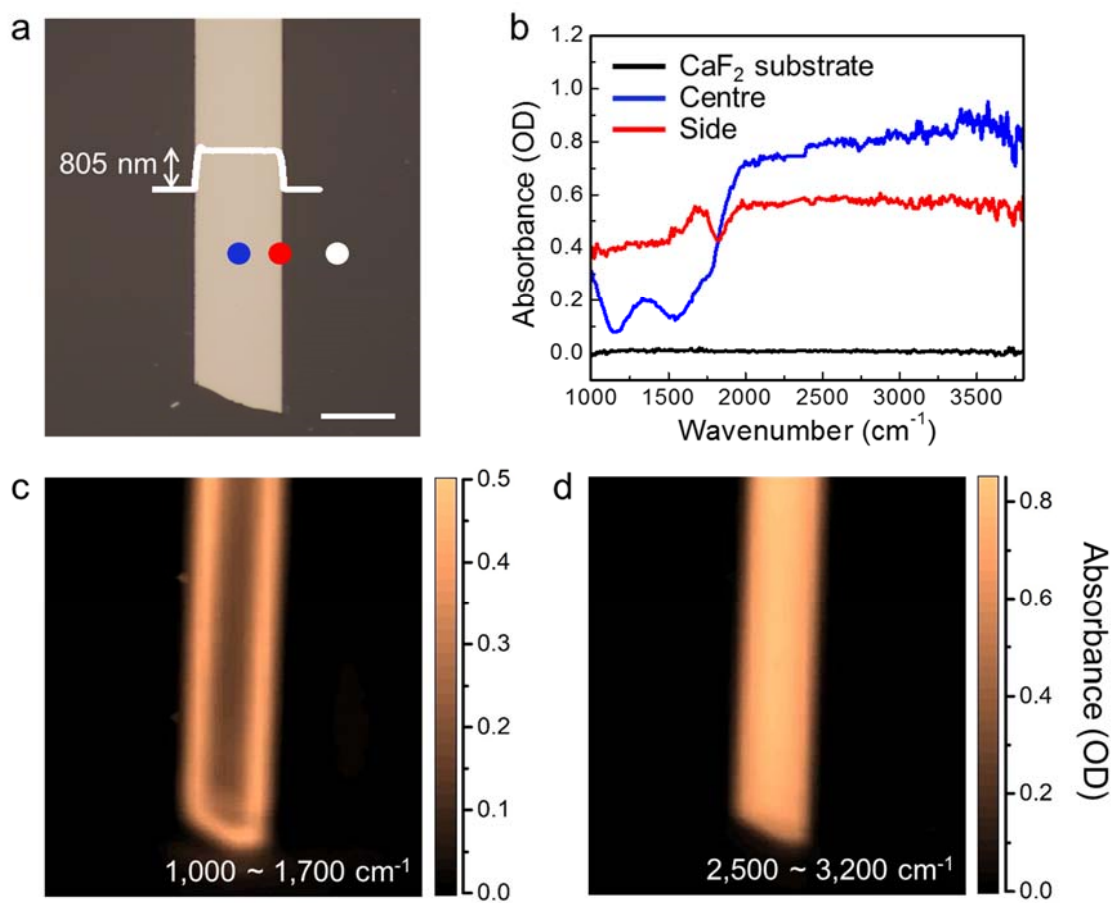

**Fig. 3 Infrared absorption of α-Bi₄Br₄ belt. (a)** OM image of the belt. Scale bar, 20 μm. **(b)** Infrared absorption spectra acquired from the belt centre (blue curve), belt side (red curve) and the CaF$_2$ substrate (black curve). The blue, red and white dots in **a** indicate the positions where the spectra are acquired. **(c, d)** 2D infrared absorption maps taken in the wavenumber regions of 1,000 ~ 1,700 cm$^{-1}$ and 2,500 ~ 3,200 cm$^{-1}$, respectively.

Considering the 1D characteristic of the TOES of α-Bi₄Br₄, the infrared absorption should be sensitive to the polarization of incident lights. For this sake, we collect infrared absorption spectra by using the linearly polarized incident light nominally perpendicular to the top (*ab* plane) of the α-Bi$_4$Br$_4$ belt with the polarization direction along the *a*- or *b*-axis, as depicted in Fig. 4a. The absorbance of the *b*-polarized light is stronger than that of the *a*-polarized light in the small wavenumber region regardless of positions of the belt, while the absorbance is nearly identical for different polarizations in the large wavenumber region. A broad peak with *b*-polarized light and a broad dip with *a*-polarized light in the region of 1,000 ~ 1,700 cm$^{-1}$ have also been observed, respectively, due to the interference of lights between the top and bottom surfaces of the belt with the thickness of 805 nm. To demonstrate the anisotropic feature of infrared absorption more clearly, we focus on the variation of the infrared absorbance vs. the polarization direction of the incident light in the small wavenumber region (1,000 ~ 1,700 cm$^{-1}$), as illustrate in Fig. 4b. The absorbance decreases dramatically with the polarization changing from the *b*-direction (0°) to the *a*-direction (90°). Furthermore, these anisotropic behaviors are also robust and can be discerned in the 2D infrared absorption maps for the *b*- and *a*-polarized incident light (Fig. 4c and d), implying that the anisotropic infrared absorption from the sides in the small wavenumber region may come from the TOES.

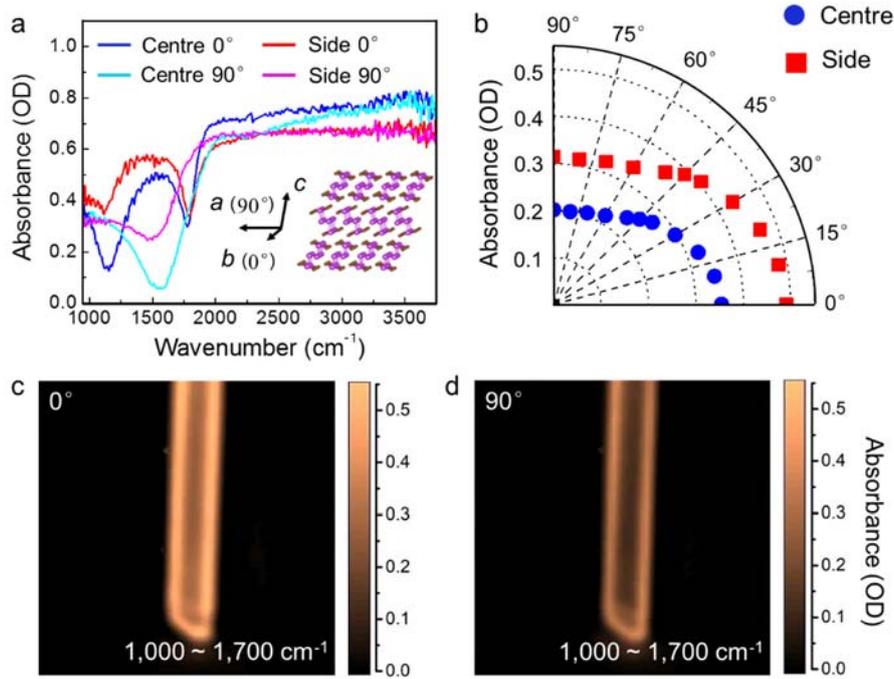

**Fig. 4 Polarization-resolved infrared absorption of α-Bi$_4$Br$_4$ belt. (a)** Infrared absorption spectra taken with linearly polarized light at different positions of the belt. The insert shows the polarization direction of the incident light with respect to the crystal lattice. **(b)** Infrared absorbance taken at the centre and side in the wavenumber region of 1,000 ~ 1,700 cm$^{-1}$ as a function of the polarization direction of the incident light. **(c, d)** 2D infrared absorption maps taken in the wavenumber region of 1,000 ~ 1,700 cm$^{-1}$ with *b*- and *a*-polarized incident lights, respectively.

To explore the physical origin of the strong infrared absorption from the belt sides and its anisotropic behavior along different polarized directions, we perform first-principles calculations on the optical response of the bulk states and TOES (for details see Supplemental materials [29]). Figure 5a presents the calculated absorption coefficient map of the *b*-polarized infrared light. The line cuts of Fig. 5a along the indicated horizontal and vertical lines are plotted in Fig. 5b and c. In line with the experimental results, the absorption from the belt sides and centre is comparable for photon energy ($E_p$) > 0.145 eV (theoretic band gap, ca. 1,170 cm$^{-1}$), while it is dominated by the belt sides for $E_p$ < 0.145 eV. Similar behaviors are disclosed in the *a*-polarized case except a weaker absorption, in particular at the belt sides for $E_p$ < 0.145 eV (Fig. 5d, e, f). The calculations reveal that the absorption of light with $E_p$ > 0.145 eV mainly arises from the inter-band excitation among the bulk states. Meanwhile, for the light with $E_p$ < 0.145 eV, two kinds of electron excitation processes, (a) from the TOES to the bulk states or vice versa and (b) between the TOES, contribute to the strong infrared absorption at the belt sides where the TOES play a key role. Moreover, due to the 1D structure of α-Bi$_4$Br$_4$, the optical transition matrix elements along the *b* direction is usually larger than that along the *a* direction, giving rise to the stronger absorption of the *b* polarized light, *i.e.* anisotropic infrared absorption, which is in good agreement with the experiments.

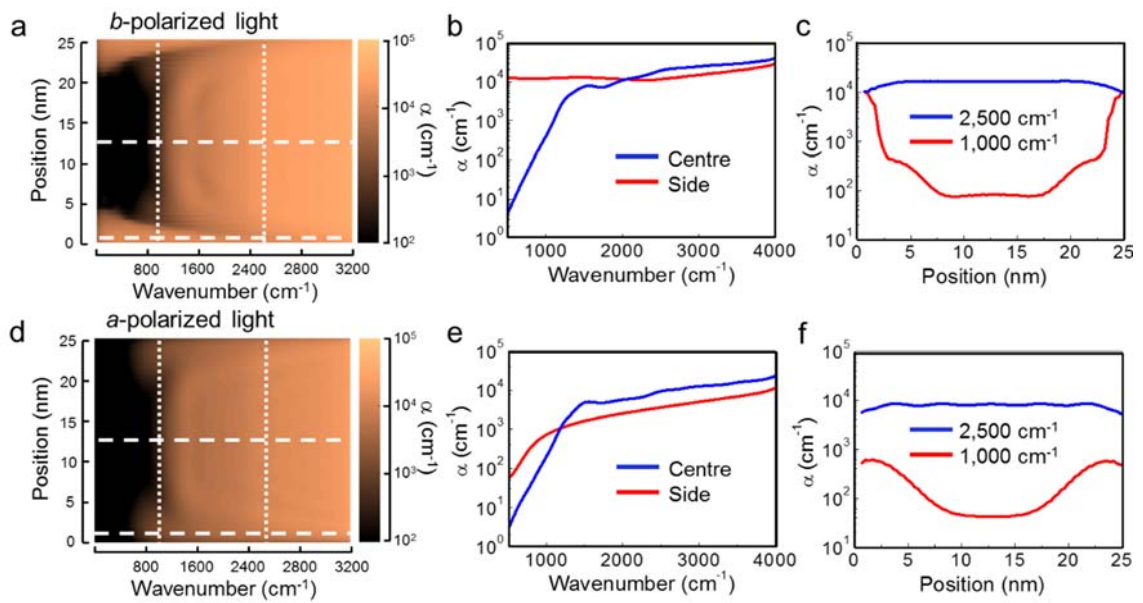

**Fig. 5 Calculated optical absorption of an α-Bi$_4$Br$_4$ belt with linearly**

**polarized incident lights. (a, d)** Spatial distribution of infrared absorption with *b*- and *a*- polarized incident light, respectively. The belt width is 25 nm and the centre is located at the position of 12.5 nm. **(b, e)** Line cuts along the horizontal lines indicated in **a** and **d**, respectively, showing the optical absorption at the centre and the side of the belt as a function of wavenumber of the polarized incident light. **(c, f)** Line cuts along the vertical lines indicated in **a** and **d**, respectively, showing the optical absorption at different positions of the belt with polarized incident light at wavenumbers of 1,000 cm$^{-1}$ and 2,500 cm$^{-1}$.

In summary, we have successfully detected the strong absorption from the TOES of quasi-1D α-Bi$_4$Br$_4$ belts using infrared spectroscopy. Under the incident light with photon energy smaller than the bulk gap, the infrared absorption from the sides of α-Bi$_4$Br$_4$ belts is much stronger than that from the belt centre and exhibits remarkable anisotropy with respected to light polarizations. These behaviors are robust and therefore assigned to the excitations involving the TOES, which is further confirmed by first-principles calculations. This work paves a new way to probing 1D edge states of topological materials. The robust TOES at room temperature and large bulk band gap makes α-Bi$_4$Br$_4$ a potential candidate for future fundamental studies and device applications.

This work is funded by the National Science Foundation of China (NSFC) (11734003), the National Key Research and Development Program of China

(2016YFA0300600 & 2016YFA0300904). W.D.X. is supported by NSFC (51661135026). Y.G.Y. is supported by NSFC (11574029) and the Strategic Priority Research Program of Chinese Academy of Sciences (XDB30000000). We are grateful to A-Xin Lu in Instrument Analysis Centre of Xi'an Jiaotong University for assistance with infrared absorption measurement.